\documentclass
[preprint,prl,superbib,superscriptaddress,byrevtex,showpacs]{revtex4}%
\usepackage{amsfonts}
\usepackage{amsmath}
\usepackage{amssymb}
\usepackage{graphicx}%
\begin{document}
\preprint{ }
\title[Short title for running header]{Spin gap behaviour in a 2-leg spin-ladder BiCu$_{2}$PO$_{6}$ }
\author{B. Koteswararao}
\affiliation{Department of Physics, Indian Institute of Technology Bombay, Mumbai 400076, India}
\author{S. Salunke}
\affiliation{Department of Physics, Indian Institute of Technology Bombay, Mumbai 400076, India}
\author{A.V.Mahajan}
\affiliation{Department of Physics, Indian Institute of Technology Bombay, Mumbai 400076, India}
\author{I. Dasgupta}
\affiliation{Department of Physics, Indian Institute of Technology Bombay, Mumbai 400076, India}
\author{J. Bobroff}
\affiliation{Laboratoire de Physique des Solides, Univ. Paris-Sud, 91405 Orsay, France.}

\begin{abstract}
We present magnetic suscceptibility and heat capacity data on a new $S=1/2$
two-leg spin ladder compound BiCu$_{2}$PO$_{6}$. \ From our susceptibility
analysis, we find that the leg coupling $J_{1}/k_{B}$ is $\sim$\ $80$ K and
the ratio of the rung to leg coupling $J_{2}/J_{1}\sim$\ $0.9$. \ We present
the magnetic contribution to the heat capacity of a two-leg ladder. The
spin-gap $\Delta/k_{B}$ $=34$ K obtained from the heat capacity agrees very
well with that obtained from the magnetic susceptibility. \ Significant
inter-ladder coupling is suggested from the susceptibility analysis. \ The
hopping integrals determined using $N$th order muffin tin orbital (NMTO) based
downfolding method lead to ratios of various exchange couplings in agreement
with our experimental data. \ Based on our band structure analysis, we find
the inter-ladder coupling in the $bc$-plane $J_{3}$ to be about $0.75J_{1}$
placing the compound presumably close to the quantum critical limit.

\end{abstract}
\volumeyear{year}
\volumenumber{number}
\issuenumber{number}
\eid{identifier}
\date[Date text]{date}
\received[Received text]{date}

\revised[Revised text]{date}

\accepted[Accepted text]{date}

\published[Published text]{date}

\startpage{1}
\endpage{ }
\maketitle

\textit{Introduction:} Following the discovery of high-temperature
superconductivity (HTSC) in the cuprates \cite{bednorz1986}, there has been an
increased focus on the properties of low-D antiferromagnetic systems. \ This
is due to the innate exotic properties of these magnetic systems themselves
and their supposed connection with HTSC. \ Significant work has taken place
recently elucidating the properties of $S$ = $1/2$ and $S$ = $1$ Heisenberg
chains and their response to impurity substitutions.\ Whereas quantum
fluctuations prevent long-range order (LRO) in 1D Heisenberg systems, 3D
systems exhibit conventional LRO. \ On the other hand, in 2D systems where the
strength of magnetic interactions and quantum fluctuations can be comparable,
one might expect competing ground states and a quantum critical point
separating them. \ Spin-ladders serve as a bridge between one-dimensional (1D)
and two-dimensional (2D) magnetic systems and it is believed that an improved
understanding of spin-ladders will lead to a better understanding of magnetism
in the 2D systems. \ A major step was taken in this direction with the
prediction of spin-gaps in even-leg ladders and their absence in odd-leg
ladders.\cite{gopalan1994} followed by experimental verification in SrCu$_{2}%
$O$_{3}$ (2-leg ladder) and Sr$_{2}$Cu$_{3}$O$_{5}$ (3-leg
ladder).\cite{azuma1994} \ However, in spite of the large experimental effort,
only a small number of gapped ladders have been synthesized and studied. \ Of
these, only two (LaCuO$_{2.5}$ and Sr$_{14}$Cu$_{24}$O$_{41}$) could be doped
significantly with holes of which only the latter becomes
superconducting\cite{dagotto1999}. \ Some other compounds which have been
investigated are (C$_{5}$H$_{12}$N)$_{2}$CuBr$_{4}$ (Ref. \cite{watson2001}),
Cu$_{2}$(C$_{5}$H$_{12}$N$_{2}$)$_{2}$Cl$_{4}$ (Ref. \cite{chaboussant1997}),
Cu$_{2}$(C$_{5}$H$_{12}$N$_{2}$)$_{2}$Br$_{4}$ (Ref. \cite{deguchi2000}) which
have substantially smaller spin-gaps. There is continued effort to synthesize
and study new low-D systems since they provide a rare opportunity to elucidate
the significance of low-dimensionality, spin-gap, etc. to HTSC as also allow
to examine impurity/doping effects in a strongly correlated cuprate. \ 

In this paper, we report on the preparation and properties of a new cuprate
which, we demonstrate, can be modeled as a two-leg ladder system with
significant inter-ladder coupling in the $bc$-plane and negligible interplanar
coupling. The spin-gap, as determined from our susceptibility and heat
capacity measurements is about $34$ K while the intraladder leg coupling is
about $80$ K. Our electronic structure calculations within the framework of
$N$th order muffin tin orbital (NMTO) downfolding method \cite{okanmto} yield
hopping integrals between various Cu atoms. Using the NMTO downfolding method,
we calculate the Wannier-like effective orbitals which illustrate the shape
and extent of the active Cu orbitals and therefore indicate the exchange
pathways which lead to the ladder topology. From a practical standpoint, the
estimated $J/k_{B}$ $\approx$ $80$ K provides a unique opportunity to examine
the excitations of the coupled ladder system at temperatures ranging from well
above $J/k_{B}$ to well below $J/k_{B}$. Impurity substitutions will then
allow to probe the nature of magnetic effects thus induced, in a wide
temperature range.

\textit{Crystal structure and measurements}: Our measurements are on single
phase, polycrystalline BiCu$_{2}$PO$_{6}$ samples (space group $Pnma$ with
lattice parameters $a=11.776$ \AA , $b=5.1776$ \AA , $c=7.7903$ \AA ).

A schematic diagram of the structure is shown \ in Fig. 1. \ The unit cell
contains four formula units, with two inequivalent Cu (Cu1 and Cu2) sites and
four inequivalent O (O1-O4) sites. The characteristic feature of the structure
are CuO$_{5}$ distorted square pyramids, with Cu$^{2+}$ ion at the centre of
the five fold oxygen co-ordination. Two such pyramids share an edge formed
from a pair of basal oxygens (O2) to give rise to a Cu dimer with an
intradimer distance of $2.8$ \AA . Along the $b$-axis, each dimer connects two
others by its four O1 corners resulting in a zigzag double chain (ladder)
running along the $b$-axis (see Fig. 1). The interdimer cohesion is further
strenghtened by PO$_{4}$ tetrahedra that connect two consecutive dimers by O2
corners. The Bi ions are positioned between two ladders. \ The Cu-O-Cu angle
along the leg is about $112^{0}$ and that along the rung is about $92^{0}$.
\ In Fig. 1, various exchange couplings ($J_{1}$, $J_{2}$, etc.) and hopping
integrals ($t_{1}$, $t_{2}$, etc.) have been indicated.

Our results of the susceptibility $\chi_{meas}$ (magnetisation $M$ divided by
applied field $H$) as a function of temperature $T$ using a vibrating sample
magnetometer (VSM) of a Physical Property Measurement System (PPMS) from
Quantum Design are plotted in Fig. 2. As seen, $\chi_{meas}$ has a broad
maximum around $57$ K (indicative of a low-D magnetic system) below which it
drops rapidly (suggestive of a spin-gap). A very weak low-temperature upturn
is seen below $6.5$ K, likely due to extrinsic paramagnetic impurities and/or
natural chain breaks in our polycrystalline sample. We now analyse these data quantitatively.

An analytical solution for the spin-susceptibility of two-leg ladders in the
full $T$-range is not known. \ However, Johnston, based on extensive quantum
Monte Carlo (QMC) simulations\cite{johnston2000} has proposed an equation
which accurately reproduces the QMC-determined susceptibilities at discrete
temperatures. \ This equation (not reproduced here since it is unwieldy) is
useful for determining the exchange couplings by fitting the measured
susceptibility data and has been used to analyse such data in the two-leg
ladder SrCu$_{2}$O$_{3}$.\ We then fit (dashed line in Fig. 2) $\chi_{meas}$
to $\chi_{o}+C/(T-\theta)+m\chi_{ladder}(T)$ where the fitting parameters are
$\chi_{o}$, $C$, $\theta$, $J_{2}/J_{1}$, $J_{1}$, and $m$. Here
$\chi_{ladder}(T)$ is the $\chi$ of isolated ladders as given by
Johnston.\cite{johnston2000} \ In the absence of a generic fitting function
which can take into account arbitrary inter-ladder interactions, we attempt to
do so using the parameter $m$. \ With $m$ as a variable, the obtained
parameters are $\chi_{o}$ $=(4.4\pm0.1)\times10^{-4}$cm$^{3}/$mol Cu, $C$ $=$
($3.0\pm0.2)\times10^{-4}$cm$^{3}$K$/$mol Cu, $\theta\sim0$ K, $J_{2}%
/J_{1}=0.87\pm0.05$, $J_{1}/k_{B}=(80\pm2)$ K, and $m=0.41\pm0.02$. \ The
value of the spin-gap using\cite{johnston2000} $\Delta/J_{1}=$ $0.4030(\frac
{J_{1}}{J_{2}})+0.0989(\frac{J_{1}}{J_{2}})^{3}$ is about $34$ K. \ The Curie
constant corresponds to less than $0.1$ \% of isolated $S=1/2$ impurities.
\ This value is comparable to typical parasitic Curie terms found in single
crystals, indicating the very high quality of our samples. Since the
core-diamagnetic susceptibility\cite{selwood} $\chi_{core}$\ is $-0.6\times
10^{-4}$cm$^{3}/$mol, $\chi_{o}-\chi_{core}$ yields the Van Vleck
susceptibility $\chi_{VV}=5\times10^{-4}$cm$^{3}/$mole which is somewhat
higher than $\chi_{VV}$ of other cuprates. \ We show in Fig. 2 the curve for
isolated 2-leg ladders (with $J_{1}/k_{B}=80$ K). We also show the simulated
curve for a uniform, 2D $S=1/2$ HAF with $J/k_{B}=80$ K where the high-$T$
behavior is generated using the series expansion given by Rushbrooke and
Wood.\cite{rushbrooke1958} Also, Johnston\cite{johnston1996} parametrized the
low-$T$ $\left(  \frac{k_{B}T}{J}\leq1\right)  $ simulations of
Takahashi\cite{takahashi1989} and Makivic and Ding,\cite{makivic1991} which we
use. \ The experimental data are lower than both, the 2D\ HAF curve and the
isolated ladder susceptibility. \ This new behavior points to the importance
of a next-near-neighbor (NNN) interaction along the leg (which might be
expected due to the zig-zag nature of the leg) which might be frustrating and
might even enhance the spin-gap. In a latter section, based on our
band-structure calculations, we actually find significant NNN as also
inter-ladder couplings. \ The absence of LRO inspite of these deviations from
the isolated ladder picture should motivate the theorists to refine their
models of such systems. \ In the inset of Fig. 2, we plot the normalised
susceptibility $\chi^{\ast}(T)$ $=$ $\chi_{spin}(T)J_{1}/(Ng^{2}\mu_{B}^{2})$
(where $\chi_{spin}(T)=\chi_{meas}-\chi_{o}-C/T$) as a function of
$k_{B}T/J_{1}$. \ We find $\chi^{\ast,\max}$ ( i.e., $\chi^{\ast}$ at the
broad maximum) to be about $0.05$ which is lower than the expected value for
isolated ladders of about $0.12$.

To further confirm the spin-gap nature of BiCu$_{2}$PO$_{6}$, we did heat
capacity $C_{p}$ measurements. Since the lattice $C_{p}$ dominates the data,
it has so far not been possible to experimentally determine the magnetic
contribution to $C_{p}$ in any spin-ladder compound unambiguously. In the
present case, we are fortunate to have a non-magnetic analog of BiCu$_{2}%
$PO$_{6}$ in BiZn$_{2}$PO$_{6}$. We have then determined the magnetic heat
capacity $C_{M}$ of BiCu$_{2}$PO$_{6}$ by subtracting the measured $C_{p}$ of
BiZn$_{2}$PO$_{6}$ from that of BiCu$_{2}$PO$_{6}$ (see Fig. 3 inset). Ours
are the first reported $C_{M}$ data in a spin-ladder compound. The data are
fit to \cite{johnston2000} $C_{M}(T)=\frac{3}{2}Nk_{B}\left(  \frac{\Delta
}{\pi\gamma}\right)  ^{1/2}\left(  \frac{\Delta}{k_{B}T}\right)  ^{3/2}\left[
1+\frac{k_{B}T}{\Delta}+0.75\left(  \frac{\Delta}{k_{B}T}\right)  ^{2}\right]
\exp(\frac{-\Delta}{k_{B}T})$ shown by the solid line (Fig. 3 inset). \ From
the fit, the spin gap $\frac{\Delta}{k_{B}}$ $\sim$ $34$ K, in excellent
agreement with our susceptibility results. \ 

\textit{First principles study}: \ The local density approximation-density
functional theory (LDA-DFT) band structure for BiCu$_{2}$PO$_{6}$ is
calculated using the linearized-muffin-tin-orbital (LMTO) method based on the
Stuttgart TB-LMTO-47 code. \cite{okalmto47} The key feature of the non
spin-polarized electronic structure presented in Fig. 4 are eight bands
crossing the Fermi level which are well separated from rest of the bands.
These bands are predominantly derived from the antibonding linear combination
of Cu \textit{d}$_{{x^{2}-y^{2}}}$ and basal O \textit{p}$_{{\sigma}}$ states
in the local reference frame where the $z$-axis is along the shortest Cu-O
bond while the $x$ and $y$ axes point along the basal oxygens O1 and O2. The
band structure is 2D with practically no dispersion perpendicular to the
ladder plane (along $\Gamma X$). The eight band complex is half-filled and
metallic as expected in LDA. It lies above the other occupied Cu-$d$, O-$p$,
and Bi-$s$ character dominated bands. The P ($s,p$) and Bi ($p$) derived
states are unoccupied and lie above the Fermi level, with the Bi-$p$ states
having non-negligible admixture with the conduction bands. This admixture of
the conduction band with Bi-$p$ states is important in mediating the Cu-Cu
inter-ladder exchange coupling. Starting from such a density functional input
we construct a low-energy model Hamiltonian using the NMTO downfolding
technique. This method \cite{okanmto} extracts energy selective Wannier-like
effective orbitals by integrating out high energy degrees of freedom. The few
orbital Hamiltonian is then constructed in the basis of these Wannier-like
effective orbitals. Here, we shall retain only Cu \textit{d}$_{{x^{2}-y^{2}}}$
orbital in the basis and downfold the rest. The effective Cu \textit{d}%
$_{{x^{2}-y^{2}}}$ muffin tin orbitals generated in the process will be
renormalised to contain in their tail other Cu-$d$, O-$p$, Bi, and P states
with weights proportional to the admixture of these states with Cu
\textit{d}$_{{x^{2}-y^{2}}}$. Fourier transform in the downfolded Cu
\textit{d}$_{{x^{2}-y^{2}}}$ basis gives the desired tight-binding
Hamiltonian
%\begin{equation}
$\mathcal{H}=\sum_{\langle i,j\rangle}t_{ij}\left(  c_{j}^{\dagger}c_{i}%
+c_{i}^{\dagger}c_{j}\right)  \label{tb hamiltonian}$
%\end{equation}
in terms of the dominant Cu-Cu hopping integrals $t_{ij}$. This tight binding
Hamiltonian will serve as the single electron part of the many-body Hubbard
model relevant for this system and can be mapped to an extended Heisenberg
model with the exchange couplings related to the LDA hoppings by
%\begin{equation}
$J_{ij}=\frac{4t_{ij}^{2}}{U_{eff}}\label{heisenberg J}$
%\end{equation}
where $U_{eff}$ is the screened onsite Coulomb interaction. The various
hoppings are displayed in Table 1 and indicated in Fig. 1. The intra-dimer
(rung) hopping proceeds mainly via the edge sharing oxygens while the
inter-dimer interaction (leg hopping) proceeds via the corner sharing oxygens
with support from the PO$_{4}$ complex. \ As anticipated in the experiments,
we do indeed find that the ratio of the rung hopping to the leg hopping
$J_{2}$/$J_{1}\approx$ $1$. We find that the NNN coupling along the leg
$J_{4}$ is about $0.3$ $J_{1}$. Depending on the relative sign of this
interaction with respect to that of $J_{1}$ one might get significant
frustration effects which should also have a bearing on the ground state of
the system. We also find an appreciable coupling between the ladders ($J_{3}%
$/$J_{1}\approx$ $0.75$) mediated primarily by the unoccupied Bi$-p$ states.
Our conclusion is further supported by the plot of the corresponding Cu
\textit{d}$_{{x^{2}-y^{2}}}$ Wannier function in Fig. 5. We find that each Cu
\textit{d}$_{{x^{2}-y^{2}}}$ orbital in the unit cell forms strong pd$\sigma$
antibonding with the neighboring O-$p_{x}$/O-$p_{y}$ orbitals resulting in the
conduction band complex. The Cu ions strongly couple along the leg as well as
the rung of the ladder confirming the hoppings in either direction should be
comparable. The O-$p_{x}$/O-$p_{y}$ tails bend towards the Bi atom, indicating
the importance of the hybridization effect from the Bi cations and therefore
enhances the Cu-Cu interladder exchange interaction (see table 1).

\textit{Conclusion}: \ In conclusion, we have presented a new $S=1/2$ 2-leg
ladder BiCu$_{2}$PO$_{6}$. \ From our $\chi$ and $C_{M}$ data we obtain a
spin-gap $\Delta/k_{B}\sim34$ K and a leg coupling $J_{1}/k_{B}\sim80$ K.
\ From our first principles LDA-DFT calculations, we find $J_{2}/J_{1}\sim1$
and a significant inter-ladder interaction in the corrugated $bc$-plane
($J_{3}/J_{1}\sim0.74$). \ Considering that the uniform $S=1/2$ 2D AF\ system
has an ordered ground state, we feel that the strong inter-ladder interaction
in BiCu$_{2}$PO$_{6}$ places it close to a quantum critical point. \ The
moderate value of the $\Delta/k_{B}$ in BiCu$_{2}$PO$_{6}$ will allow one to
explore the magnetic properties in a large $T$ range, well below and well
above the gap temperature, enabling a comparison with and refinement of
theoretical models. \ \ We feel that there might still be unanticipated
features in the physics of low-D magnets and we expect our work to motivate
others to carry out numerical simulations and explore the phase-diagram of
coupled 2-leg ladders in the presence of NNN couplings along the leg. \ We are
presently considering doping/substitutions in this 2-leg ladder which might be
able to tune the inter-ladder exchange and effect a quantum phase transition.

\begin{acknowledgments}
\ We thank the Indo-French Center for the Promotion of Advanced Research for
financial support.
\end{acknowledgments}

\textbf{Table Caption}

Table 1 Hopping parameters ($t_{n}$) between various Cu's are tabulated along
with the corresponding Cu-Cu distances. \ The hopping paths are indicated in
Fig. 1.%

\begin{tabular}
[c]{|c|c|c|c|}\hline
Hopping path & Cu-Cu distance & $t_{n}$ & $J_{n}/J_{1}$\\\hline
& in \AA  & in meV & $=(t_{n}/t_{1})^{2}$\\\hline
leg ($t_{1}$) & 3.22 & 155 & 1\\\hline
rung ($t_{2}$) & 2.90 & 154 & 1\\\hline
inter-ladder ($t_{3}$) & 4.91 & 133 & 0.74\\\hline
NNN in leg ($t_{4}$) & 5.18 & 91 & 0.34\\\hline
diagonal ($t_{5A}$) & 4.43 & 30 & 0.04\\\hline
diagonal ($t_{6A}$) & 5.81 & 26 & 0.03\\\hline
\end{tabular}

Figure captions

Fig. 1 (Color Online) A schematic of the BiCu$_{2}$PO$_{6}$ crystal structure
is shown. \ It can be seen that 2-leg ladders run along the crystallographic
$b$-direction. \ The 2-leg ladder is separately shown for clarity. \ Also
shown are the various significant hopping parameters/exchange couplings
between Cu atoms.

Fig. 2 (Color Online) Magnetic susceptibility ($\chi_{meas}=M/H$) vs.
temperature $T$ \ for BiCu$_{2}$PO$_{6}$ in an applied field of $5$ kG. The
open circles represent the raw data and the dashed line is a fit (see text).
Also shown are simulated curves for the isolated ladder (dark gray line) and
for the 2D\ HAF (gray line). \ The inset shows the dependence of $\chi^{\ast}$
on $k_{B}T/J_{1}$ (see text).

Fig. 3 (Color Online) The measured heat capacity as a function of $T$ for
BiCu$_{2}$PO$_{6}$ and BiZn$_{2}$PO$_{6}$. \ Inset: the magnetic specific heat
of BiCu$_{2}$PO$_{6}$ along with a fit (see text).

Fig. 4 (Color Online) LDA band dispersion of BiCu$_{2}$PO$_{6}$ along various
symmetry directions.

Fig. 5 (Color Online) Effective Cu1 $d_{{x^{2}-y^{2}}}$ orbital with lobes of
opposite signs colored as black and white. The $d_{{x^{2}-y^{2}}}$ orbital is
defined with the choice of local coordinate system as discussed in the text
(height of the isosurface $=\pm0.09$). The spheres represent the ions.

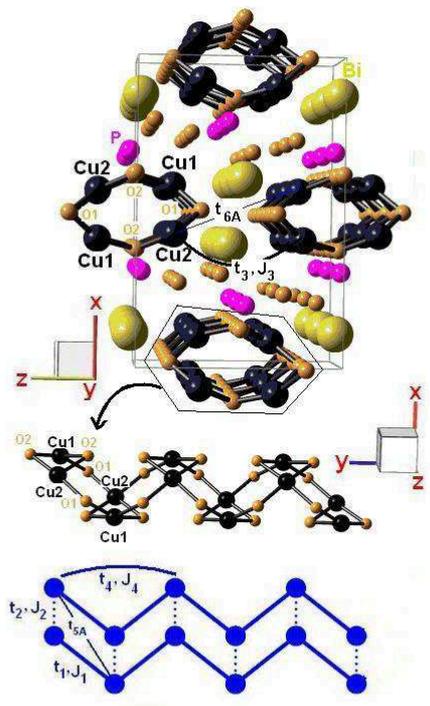
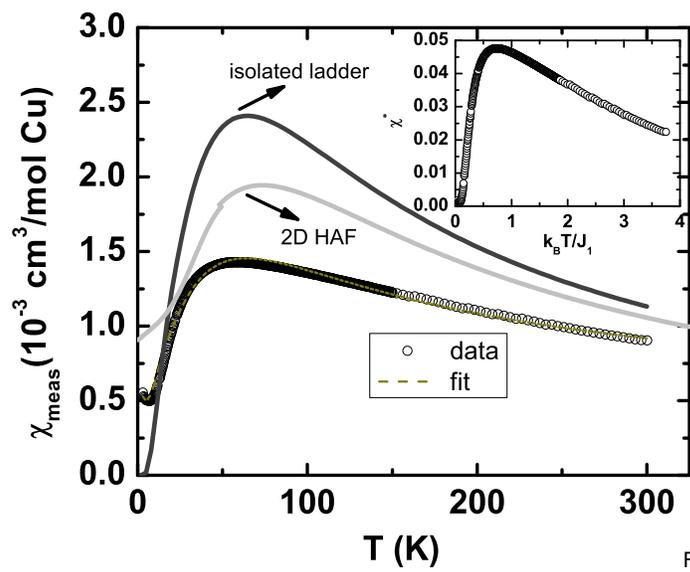
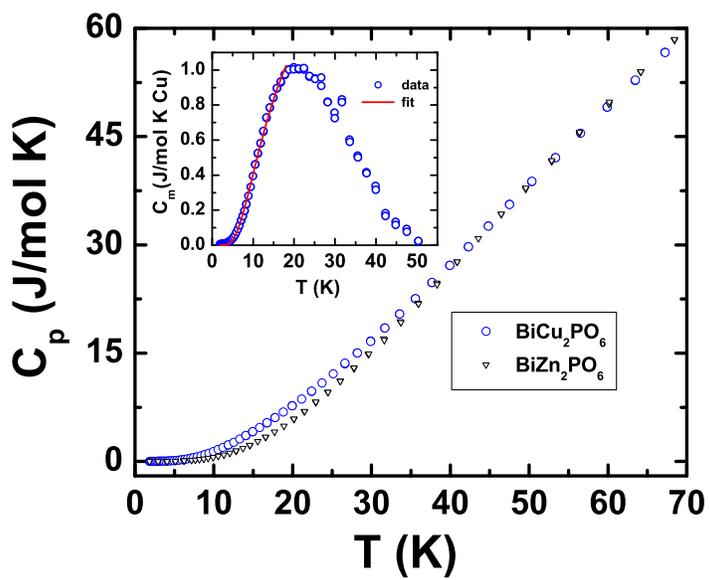
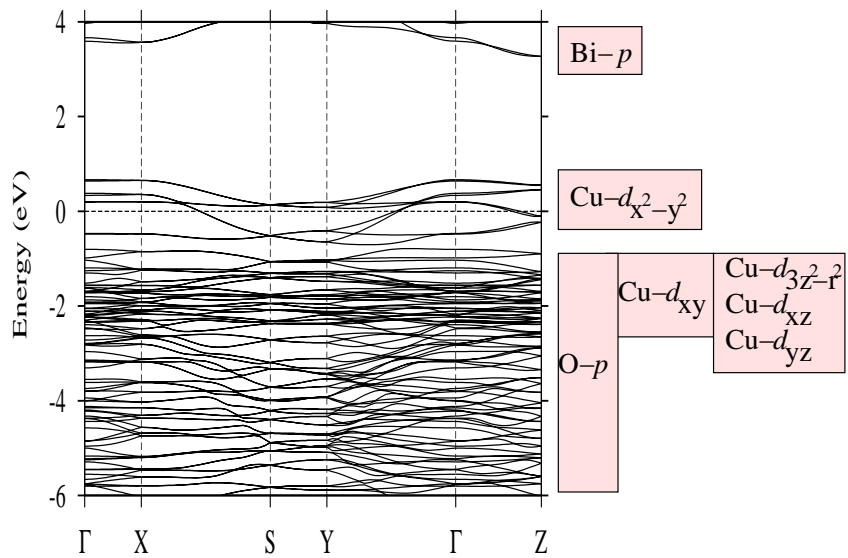
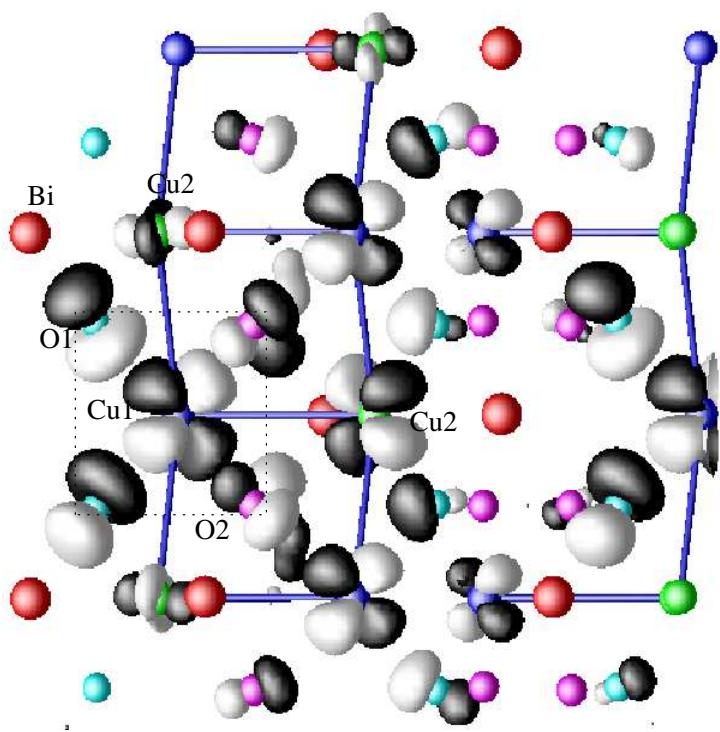
\end{document}